# La$_2$O$_3$Mn$_2$Se$_2$: a correlated insulating layered *d*-wave altermagnet

Chao-Chun Wei,[1] Xiaoyin Li,[1] Sabrina Hatt,[2] Xudong Huai,[3] Jue Liu,[4] Birender Singh,[5] Kyung-Mo Kim,[5] Rafael M. Fernandes,[6] Paul Cardon,[1] Liuyan Zhao,[7] Thao T. Tran,[3] Benjamin M. Frandsen,[2] Kenneth S. Burch,[5] Feng Liu,[1] Huiwen Ji[1,*]

[1]*Department of Materials Science and Engineering, University of Utah, Salt Lake City, Utah 84112, United States*

[2]*Department of Physics and Astronomy, Brigham Young University, Provo, Utah 84602, United States*

[3]*Department of Chemistry, Clemson University, Clemson, South Carolina 29634, United States*

[4]*Neutron Sciences Division, Oak Ridge National Laboratory, Oak Ridge, Tennessee 37831, United States*

[5]*Department of Physics, Boston College, Chestnut Hill, Massachusetts 02467, United States*

[6]*Department of Physics, University of Illinois Urbana-Champaign, Urbana, IL 61801, United States*

[7]*Department of Physics, University of Michigan, Ann Arbor, MI 48109, United States*

**ABSTRACT**

Altermagnets represent a new class of magnetic phases without net magnetization that are invariant under a combination of rotation and time reversal. Unlike conventional collinear antiferromagnets (AFM), altermagnets could lead to new correlated states and important material properties deriving from their non-relativistic spin-split band structure. Indeed, they are the magnetic analogue of unconventional superconductors and can yield spin polarized electrical currents in the absence of external magnetic fields, making them promising candidates for next-generation spintronics. Here, we report altermagnetism in the correlated insulator, magnetically-ordered tetragonal oxychalcogenide, La$_2$O$_3$Mn$_2$Se$_2$. Symmetry analysis reveals a $d_{x^2-y^2}$-wave type spin momentum locking, which is supported by density functional theory (DFT) calculations. Magnetic measurements confirm the AFM transition below ~166 K while neutron pair distribution function analysis reveals a 2D short-range magnetic order that persists above the Néel temperature. Single crystals are grown and characterized using X-ray diffraction, optical and electron microscopy, and microRaman spectroscopy to confirm the crystal structure, stoichiometry, and uniformity.

# I. INTRODUCTION

Magnetic memory and spintronics have historically relied on ferromagnetism (FM) for straightforward implementations in devices [1,2]. Over the past decade, several paradigms of antiferromagnetic (AFM) spintronics were demonstrated with promising advantages such as zero stray fields, robustness against external perturbations, potential for miniaturization, and ultrafast switching dynamics [3-6]. However, the spin degeneracy enforced throughout the band structure and the need to couple the magnetic response to electric fields limited the potential applicability of most AFM materials. To overcome this, the community focused on materials with strong spin-orbit coupling (SOC) [7-9], requiring heavy elements and complex magnetic orders that limit widespread adoption.

Recent progress in the theoretical classification of spin groups, where SOC is absent, identified a new class of magnetic states called altermagnets, which exhibit *k*-dependent spin polarization yet zero overall magnetization [10,11]. Such a non-relativistic spin splitting is a consequence of the fact that the magnetic configuration is invariant under a combination of time reversal, which flips the magnetic moments, and real-space rotations, which can be proper, improper, or accompanied by half-translations (i.e., a non-symmorphic operation) [12]. Altermagnets are suitable for making compact and high-speed spintronics and the material choices are no longer limited to heavy-metal compounds since the spin-splitting is not enforced by SOC [10,13,14]. Emerging properties such as anomalous Hall effect [11,15,16], current polarization, tunneling and giant magnetoresistance are predicted based on the symmetries of the spin density on the crystalline structure [17]. Altermagnets are divided into *d*-, *g*-, and *i*-wave types, based on the number of nodal planes (*k*-space planes on which bands are spin-degenerate) between spin-split bands in momentum space [10,18]. As such these materials form a long-sought magnetic analogue to the higher momentum, unconventional superconducting states known for decades [19]. So far, a large number of materials have been predicted to be altermagnetic [20-23], and direct experimental evidence has been reported in *g*-wave type MnTe and CrSb [24,25], while the status of $RuO_2$ (a predicted *d*-wave type altermagnet) remains unsettled [26-30].

An interesting open problem is the interplay between altermagnetism and strongly-correlated materials, which can be investigated in Mott insulating or correlated metallic materials whose crystalline symmetries "convert" the common antiferromagnetic ground state into an altermagnet. To identify new types of correlated material candidates with *d*-wave altermagnetism, we turn to the layered $RE_2O_3TM_2Ch_2$ (RE = rare earth, TM = transition metal, Ch = chalcogen) family [31-34]. This family was studied in the last decades for its structural resemblances to the layered $ThCr_2Si_2$ family (122-type) (Figure 1a) [35-37], in which Fe-based superconductors were discovered

(for a recent review, see Ref. [38]). The crystal structure of these compounds (Figure 1b) adopts 2D square arrays of TMs, such as Fe, Mn, Co that are octahedrally coordinated by 4 chalcogen and 2 O atoms in a face-sharing network. The resultant 4-fold rotation symmetry connecting the two TM sublattices provides an ideal platform to realize *d*-wave type altermagnets. Unlike most other altermagnetic candidates, rotation relating the time-reversed spin states involves two different anion species, i.e., O, Se. These materials are predominantly magnetically ordered with electrically insulating behavior regardless of the TM species, which has been attributed to a Mott insulating or orbital-selective Mott behavior [32,33,39,40]. Indeed, resistivity measurements find insulating behavior already in the paramagnetic phase, whereas DFT calculations generally predict a metallic phase [60]. In addition, the mixed anion coordination environments around TMs can lower local crystal symmetry, enhance local anisotropy [41], and, with specific AFM order, might induce altermagnetism.

In this work, we revisit this structure family and report *d*-wave type altermagnetism predicted in the layered compound $La_2O_3Mn_2Se_2$ (LOMSO). The compound was initially discovered by Ni et al and had its magnetic structure identified through powder neutron diffraction [42]. Here, symmetry analysis based on its reported G-type ***k*** = (0,0, 0) magnetic ground-state order indicates the breaking of certain symmetries and the subsequent rise of *d*-wave spin momentum locking. The prediction is supported by density-functional theory (DFT) calculations, which further show that adding SOC modifies the electronic structure by lifting spin degeneracy at nodal planes and converting them into symmetry-protected nodal lines. Magnetic measurements confirm the AFM ground state and a competing FM component. Neutron pair distribution function (PDF) analysis reveals a 2D short-range intraplanar AFM order that persists even above the magnetic ordering temperature. Finally, single crystal samples of several hundred micrometers were grown and their stoichiometry, uniformity and layered crystal structure were confirmed through single crystal X-ray diffraction, optical and electron microscopy, and microRaman spectroscopy.

## II. METHODS

La$_2$O$_3$ was prepared from heating La(OH)$_3$ (Sigma Aldrich, 99.9% trace metals basis) at 900°C for 48 hours. Manganese powder (Thermo Scientific, 99.95%), selenium shot (Thermo Scientific, 99.999%), and La$_2$O$_3$ were then combined in a stoichiometric molar ratio of 2:2:1, ground, pelletized, and sealed in a quartz tube. The mixture was first heated to 700°C, followed by regrinding, and a second heat treatment at 950°C for 48 hours. To prepare single crystals, the synthesized LOMSO powder was pelletized and placed in a high-temperature tube furnace with an argon flow. The temperature was increased at a rate of 3°C/min to 1000°C, then at 1°C/min to 1200°C, held for 24 hours. The furnace was then cooled to 1000°C at a rate of 1°C/min over 18 hours, followed by cooling to room temperature at 3°C/min.

A STOE STADI P X-ray diffractometer was used in transmission geometry to collect powder diffraction data with a Mo anode with K$\alpha$1 radiation ($\lambda = 0.7094$ Å), a Ge (111) curved monochromator, and a Dectris MYTHEN2 detector. Data was collected at room temperature in static mode across the 2$\theta$ angle range of 2° to 60° on LOMSO powder over 11.5 hours. Structural refinement was done with the GSAS-II software package. VESTA was used to visualize crystal structures.

Neutron pair distribution function (PDF) measurements were performed on the NOMAD instrument at the Spallation Neutron Source. The PDF method involves Fourier transforming the total scattering pattern (including both Bragg scattering from long-range-ordered correlations and diffuse scattering from short-range correlations) to reveal the local structure in real space [43]. When PDF experiments are performed on magnetic materials using a neutron beam, the magnetic correlations present in the material produce the magnetic PDF (mPDF) as an additional component of the total PDF data [44,45], providing insight into both the atomic and magnetic correlations on local length scales. In this work, the sample was contained in a 3 mm diameter thin quartz capillary, and an argon cryoblower was used to control the sample temperature. Data were collected at 100 K, 200 K, 300 K, and 400 K. Two 24 min scans were collected for each temperature point and summed together to improve the statistics. Neutron scattering signals were background subtracted, and the obtained intensity was normalized to the scattering signal from a 6 mm vanadium rod to correct for detector efficiency. Using the standard protocols at the beamline, the raw diffraction data were reduced to the total scattering structure function, which was then transformed into the PDF with a maximum momentum transfer of 30 Å$^{-1}$. Combined magnetic and atomic PDF co-refinements were performed using the diffpy.mpdf and diffpy.srfit packages [46,47], part of the DiffPy suite of Python-based diffraction software.

Scanning electron microscopy (SEM) (FEI TENEO) was used for imaging with a secondary electron detector and a 20 kV high voltage beam, at a magnification of 1200× and a working distance of 10.5 mm. Energy Dispersive Spectroscopy (EDS) uses EDAX Octane Elect SDD. The working distance and voltage are the same as SEM. KEYENCE VHX-500 was used for optical microscopy (OM) imaging.

Temperature- and field-dependent magnetization measurements were performed in a Quantum-Design Physical Properties Measurement System (PPMS) using a vibrating-sample magnetometer option. Magnetic susceptibility is approximated by normalizing magnetization by the field applied.

First-principles calculations of LOMSO were performed based on density-functional theory (DFT) as implemented in the Vienna Ab initio Simulation Package (VASP) [48]. The projector augmented wave method was used to describe the interactions between core-valence electrons [49], and the Perdew-Burke-Ernzerhof-type generalized gradient approximation was employed for the exchange-correlation functional [50]. The energy cutoff for the plane wave expansion was set to 520 eV, and the Brillouin zone was sampled by a $13 \times 13 \times 3$ Monkhorst-Pack mesh [51]. To describe the strongly correlated $3d$ electrons of Mn, the rotationally invariant DFT+U scheme was implemented [52], using an effective U value of 4 eV based on a similar choice made for MnO [53]. The lattice constants and atomic positions were fully relaxed to obtain the computationally optimized structure. The obtained optimal lattice constants are $a = b = 4.16$ Å, $c = 19.17$ Å, and $\alpha = \beta = \gamma = 90°$, and are reasonably close to the experimental values. Convergence criteria for energy and force were set to $10^{-5}$ eV and $-0.001$ eV/Å, respectively.

Micro-Raman spectroscopy was performed using a custom-built system with a 532nm excitation laser and 1μm spot size [54,55]. The incident and scattered polarization were controlled using motorized half-waveplates with spectra collected every fifteen degrees from 0 to 360. The incident power was kept below 0.1 mW, and we confirmed the absence of heating by doubling the integration power and reducing the power by half did not affect the spectra. A total of 10 different single crystals were measured, as well as spectra acquired from multiple spots on three different crystals. The spectra were all identical, and the polarization dependence was also highly consistent across crystals.

## III. RESULT AND DISCUSSION

Figure 1b shows the crystal structure of $RE_2O_3TM_2Se_2$. The ($TM_2Se_2O$) octahedra layers have a square pattern and are separated by the ($RE_2O_2$) spacer layers. Across a ($RE_2O_2$) spacer layer, the two nearby $TM_2Se_2O$ layers are offset in plane by $½\ \boldsymbol{a} + ½\ \boldsymbol{b}$. In this family, the presence or absence of altermagnetism depends on the magnetic configuration, which in turn depends on the TM. For example, the magnetic structure of the more widely studied $La_2O_3Fe_2Se_2$ adopts in-plane moments and a zig-zag type AFM order that doubles the unit cell along *a* and *c* axes, giving rise to a non-zero propagation vector ***k*** = (½,0, ½), as shown in Figure 1d [56]. The magnetic structure thus hosts $\tau T$ symmetry (where $\tau$ is translation and $T$ is time-reversal) and does not qualify as an altermagnet, being instead an AFM. The Co analogue, $La_2Co_2O_3Se_2$ (Figure 1e), develops a co-planar non-collinear magnetic structure below ~217 K, wherein nearby Co moments are orthogonal with a non-zero ***k*** = (½,½,0) [57]. The magnetic structure thus preserves both $PT$ (one inversion center highlighted in the lower panel of Fig. 1e) and $\tau T$ symmetry and does not lead to altermagnetism.

The Mn variant, which is relatively less explored with only one report on its magnetic structure through neutron diffraction, on the other hand, might be altermagnetic based on symmetry analysis. The G-type order as shown in Figure 1c for $La_2O_3Mn_2Se_2$ (LOMSO) [42], features opposite spins in the nearest-neighbor TMs and a checkerboard-type magnetic moment pattern that does not break translational symmetry. Two key criteria for altermagnet identification are met here: (i) The absence of parity-time reversal ($PT$) symmetry [10]. This is evident in the crystal structure, where pairs of atoms connected by inversion symmetry (marked by dashed lines) exhibit parallel magnetic moments, indicating the breaking of $PT$ symmetry. (ii) The lack of translation-time reversal ($\tau T$). The magnetic propagation vector ***k*** = (0,0,0) suggested from previous neutron scattering rules out a magnetic supercell, automatically ensuring the absence of $\tau T$ symmetry. The two sublattices are instead connected by 4-fold rotation, which is the defining property of altermagnetism. Even in the absence of SOC, a highly desired *d*-wave type spin-momentum locking is predicted with two orthogonal spin-degenerate nodal planes parallel to the 4-fold rotation axis (*c* axis).

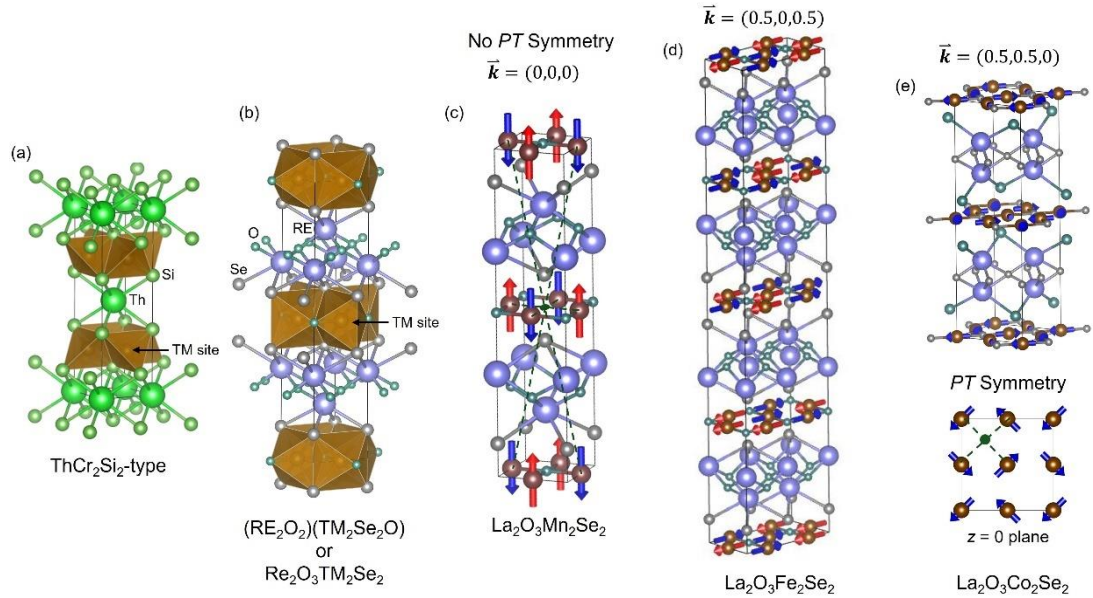

**Figure 1.** Atomic crystal structures of (a) ThCr$_2$Si$_2$ and (b) (RE$_2$O$_2$)(TM$_2$Se$_2$O) and magnetic structures of (c) La$_2$O$_3$Mn$_2$Se$_2$, (d) La$_2$O$_3$Fe$_2$Se$_2$, and (e) La$_2$O$_3$Co$_2$Se$_2$. Blue/red arrows in (c-e) point at the directions of the compensating moments while the inversion centers are highlighted by green-colored dots in (c) and (e). Representative pairs of atoms connected by inversion are connected by dashed lines in (c) and (e).

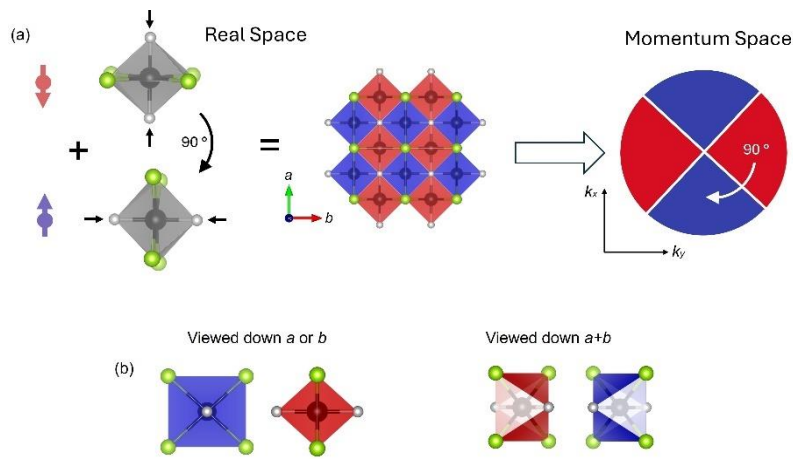

**Figure 2.** Origin of $d$-wave altermagnetism in LOMSO. (a) The 4-fold or 90° rotation between the opposite-spin sublattices (in red and blue) in real space yields $d$-wave type spin-momentum locking in momentum space. The anisotropy directions in MnSe$_4$O$_2$ octahedra are highlighted. The blue and red blocks in momentum space indicate spin-momentum locking without SOC while the spin-degenerate nodal planes are depicted using white lines. (b) Inspection of the Mn$_2$Se$_2$O octahedra down $a$ (or $b$) and $a+b$ directions sheds light upon the spin-splitting direction.

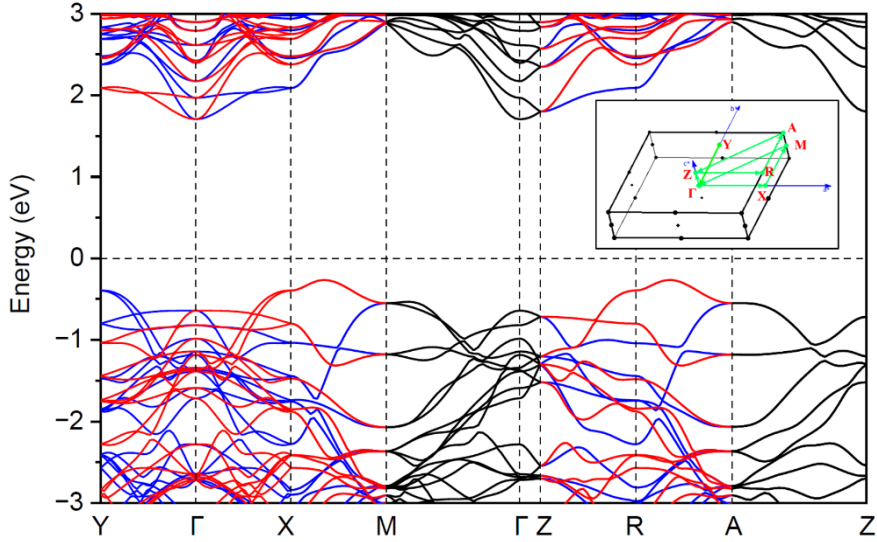

**Figure 3.** Spin-projected band structure of La$_2$O$_3$Mn$_2$Se$_2$ without SOC. Red and blue lines represent the spin-up and down polarized bands. The degenerate bands with compensated spin components are plotted as black solid lines. The first Brillouin zone is included as an inset.

The symmetry properties of the magnetically ordered state of LOMSO imply that the altermagnetic order parameter has a $d_{x^2-y^2}$-wave character, i.e., the momentum-space spin-density behaves as $S_k \propto (k_x^2 - k_y^2)\sigma$. As shown in Figure 2a, the mixed anion coordination environments create local anisotropy, i.e., two shorter Mn-O bonds and four long Mn-Se bonds. The two nearby MnSe$_4$O$_2$ octahedra are related by a 4-fold or 90° rotation and have opposite moments. Because the two Mn atoms in each unit cell are located at (1/2, 0, 0) and (0, 1/2, 0), the spin splitting is maximized along the main axes and vanishes along the diagonals, yielding the $d_{x^2-y^2}$-wave form factor. This can also be seen by comparing the two opposite-spin MnSe$_4$O$_2$ octahedra from a view down the $a$ or $b$ axis (Figure 2b), where they look different, and from a view down the $a+b$ axis (Figure 2c), where they look identical. Thus, accordingly, without SOC, the bands with opposite spins are split except along the two vertical nodal planes $k_x = \pm k_y$ (i.e., parallel to the $\Gamma - M$ directions), as shown in Figure 2a on the right. Note that this symmetry is distinct from the more widely discussed $d$-wave altermagnetic rutiles, which have a $d_{xy}$-wave character with nodal planes along $k_x = 0$ and $k_y = 0$ (i.e., parallel to the $\Gamma - X$ and $\Gamma - Y$ directions).

To confirm the above symmetry analysis, a band structure calculation is carried out and shown in Figure 3. The blue and red dispersions refer to the spin-up and down polarized bands while the black-colored bands are spin-degenerate. The first Brillouin zone of the magnetic structure is shown on the right. SOC was first purposely left out because altermagnetism is not a relativistic effect and does not rely on SOC to emerge. In the absence of SOC, LOMSO indeed exhibits a momentum-dependent spin splitting, which implies an altermagnetic order. This behavior contrasts with that of conventional FMs, where spin splitting is typically uniform across the entire Brillouin zone, and collinear AFMs, which show no net spin splitting due to the compensating magnetic moments. Significant splitting as large as nearly 0.7 eV is observed between the spin-up and spin-down channels along the $\Gamma - X(Y)$, and reverses its polarization when rotating from $\Gamma - X$ to $\Gamma - Y$, confirming our prediction of a $d_{x^2-y^2}$-wave modulation of the spin splitting. A gap >1.5 eV is also observed, which is consistent with the insulating nature generally assigned to this class of materials as well as the green color of the as-synthesized powder. The gap size is affected by the effective U parameter: a larger U leads to a larger gap (Supplementary Figure S1). Based on DFT calculations, unlike the Fe analogue, LOMSO displays a G-type compensated magnetic state even in the limit of U=0. Indeed, as shown in Supplementary Table 1, regardless of the functional used, the compensated magnetic state is always the ground state and much lower in energy compared to the FM or the non-magnetic state. While the band structure is gapped by the magnetic order even when U is set to zero, the non-magnetic state displays a metallic electronic structure (Supplementary Figure S2). Previous resistivity measurements above the magnetic transition found insulating behavior up to the highest temperatures probed [60], pointing out that this material, like the other TM members of the family, is likely a Mott insulator.

The introduction of SOC modifies the electronic structure, particularly by lifting the spin degeneracy at the nodal planes in the Brillouin zone, which without SOC are parallel to $\Gamma - M$ and $\Gamma - Z$. In the presence of SOC, the direction of the magnetic moments matters [58], impacting the momentum-dependent spin-density which, without SOC, is given by $\boldsymbol{S_k} \propto (k_x^2 - k_y^2)\boldsymbol{\sigma}$. When the moments are oriented along the $c$-axis, which is the case for LOMSO according to Ref [42] and as our neutron experiment also confirmed below, the altermagnetic order parameter transforms as the $B_{2g}^-$ irreducible representation of the tetragonal point group D$_{4h}$, following the notation of Ref [59] (the minus superscript indicates that the quantity is odd under time reversal). Using the result of Ref. [59], it follows that SOC will lift the spin degeneracy everywhere on the nodal planes except along three $\boldsymbol{k}$ lines: $k_x = k_y$, $k_z = 0$; $k_x = -k_y$, $k_z = 0$; $k_x = k_y = 0$, (i.e., along $k_z$). The prediction is verified by DFT calculations: as shown in Figures 4c,d, along $\Gamma$-A, the presence of SOC induces spin splitting, whereas spin degeneracy is retained along $\Gamma - M$ and $\Gamma - Z$ (Figures 4a,b). Interestingly, the spin-splitting nodal lines are protected by crystalline mirror symmetries [59]. A main consequence is that a magnetic field perpendicular to one mirror will not gap out the nodal lines immediately.

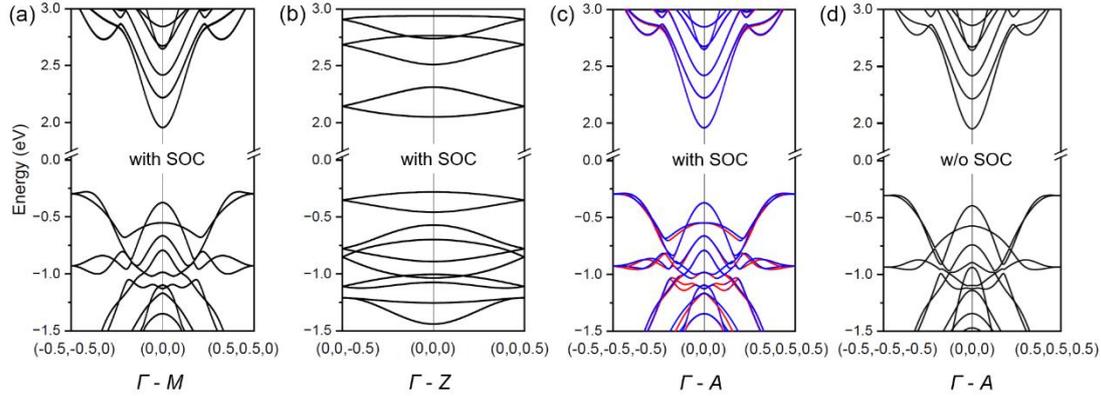

**Figure 4.** DFT electronic structure calculations (a-c) with SOC included along the $\Gamma - M$, $\Gamma - Z$ and $\Gamma$-A path, and without SOC along $\Gamma$-A. The blue and red bands are spin split while the black ones are spin-degenerate.

The target compound was synthesized through a traditional solid-state calcination method. Powder X-ray diffraction was measured, and the results of Rietveld refinement are shown in Figure 5a. A good fit (weighted R-value of 5.7%) was obtained between the observed data and the structural model with an I4/mmm space group as reported in Ref [42,60], affirming the high purity of the sample. The compensated magnetic order in LOMSO powder was then confirmed by magnetic measurements. DC magnetic susceptibility under zero-field cooling (ZFC) and field cooling (FC) at 1000 Oe was measured from 300 K down to 2 K (Figure 5b). The magnetic transition at ~166 K is indicated by a kink in the ZFC data and its divergence from that of FC. A small FM component was also observed below 140 K. The field-dependent ZFC and FC susceptibility shown in Figure 5c suggests the increasing dominance of the compensated magnetic state by an external field. The more significant upturn below ~140 K was also previously seen for the compound at ~120 K and was attributed to the onset of spin reorientation [42]. Magnetization vs. field at various temperatures is shown in Figure 5d. The magnetization stays mostly linear with the external field, consistent with an overall altermagnetic order. A small hysteresis loop develops as the temperature decreases to 2 K, indicating contribution from the FM component albeit with very small magnetic moment per mole of Mn and that the altermagnetic order still dominates.

The origin of the competing FM component, and particularly whether it is intrinsic or extrinsic, remains unclear. For instance, the competing FM component co-existing with the dominant AFM order might be explained by the Goodenough-Kanamori Rule as pointed out in Ref [42], in that the close-to-90° Mn-Se-Mn ($J_1$) and the 180° Mn-O-Mn super-exchange ($J_2$) interactions between next-nearest neighbors should be FM-like and AFM-like, respectively, which potentially give rise to magnetic frustration (Figure

5f). The nearest neighbor Mn-Mn direct exchange interaction $J_3$ should be AFM-like as inferred from the G-type ground state observed. In contrast to this intrinsic scenario, the FM component below 140 K was attributed by one report [60] to small amounts of $LaMnO_{3+\delta}$ that order ferromagnetically with divergence between FC and ZFC data below ~140 K [61]. But the FM moment as observed in $LaMnO_{3+\delta}$ was enhanced by an increasing external field rather than being suppressed as we saw in our measurements.

Magnetic short-range order was previously reported based on a broad bump in neutron diffraction that corresponds to 2D magnetic order and does not completely disappear even at 300 K [42]. We therefore collected additional susceptibility data between 300 – 400 K for a Curie-Weiss fit, which is plotted in Figure 5e. The significant negative $\theta_{CW}$ = –1684(3) K further indicates strong AFM-like interactions between nearest-neighbors, and the effective magnetic moment of 5.31 $\mu_B$/Mn is reasonably close to the theoretical value for high-spin $Mn^{2+}$, $\sqrt{S(S+1)}$ = 5.92 $\mu_B$/Mn. We note that given the octahedral coordination, we would expect the T2g level of the Mn to be partially filled (i.e., $d^5$ state), thus the strong insulating nature and high spin state suggest a correlated insulator of likely Mott origin.

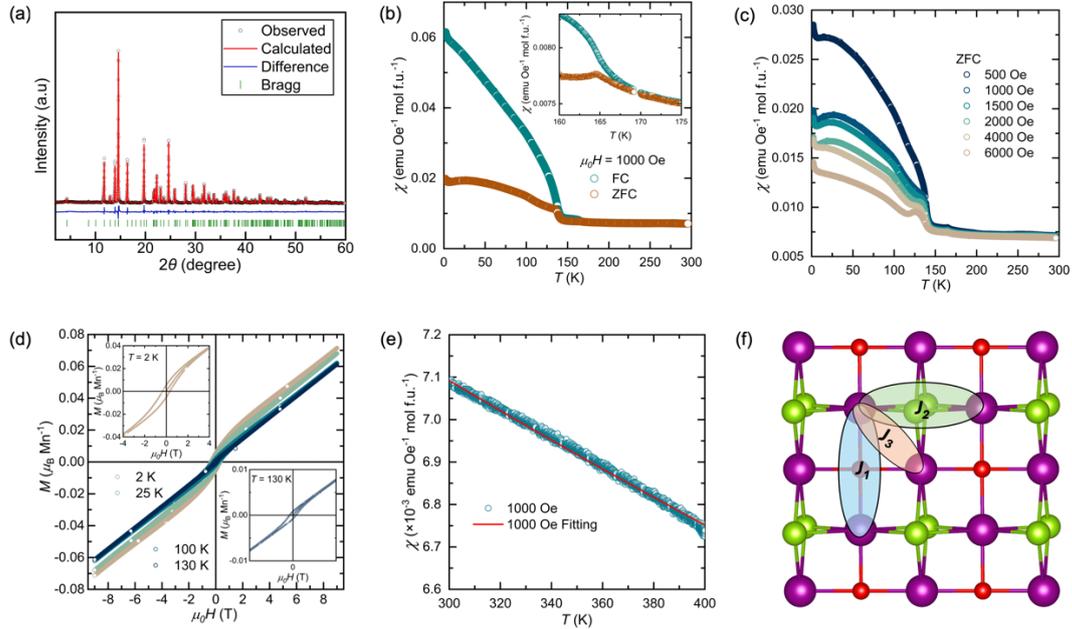

**Figure 5.** Characterizations of LOMSO powder. (a) Rietveld refinement of X-ray powder diffraction pattern. (b) DC ZFC and FC magnetic susceptibility in an external magnetic field of 1000 Oe. The transition is highlighted in the zoom-in panel. (c) Magnetic susceptibility (ZFC) at various fields. (d) Magnetization vs. field at various temperatures. (e) Curie-Weiss fit in a high-temperature range. (f) Schematic of co-existing exchange interactions $J_1$, $J_2$, and $J_3$ in the $Mn_2Se_2O$ layer, wherein the purple, green, and red balls represent Mn, Se, and O, respectively.

A previous neutron diffraction study found an asymmetric broad bump underneath the magnetic (101) peak between 100 and 300 K and tentatively attributed it to possible 2D local order [42]. To better characterize the local magnetic interactions, we used a PDF method, which uses a Fourier transform of the diffraction data to highlight real-space correlations, i.e., short-range order. Neutron PDF data was collected using a time-of-flight neutron source. The PDF of LOMSO at 100 K is shown in Fig. 6a, together with a fit to the data using a model of the atomic structure only (i.e., no magnetic component was included in the fit shown). The positions and heights of the sharp peaks in the data are well described by the fit, confirming that the accepted tetragonal crystal structure for LOMSO provides a good starting point for the PDF data. However, an inspection of the fit residual (green curve offset vertically below) reveals somewhat longer-wavelength modulations, as well as a positive bump below about 1.5 Å. These features are indicative of a magnetic component to the total PDF signal, as expected from the known long-range magnetic order that forms below 166 K in this system.

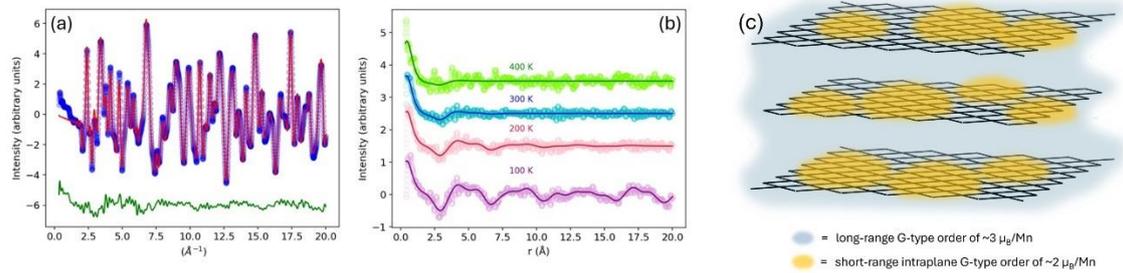

**Figure 6.** (a) Atomic PDF fit to the LOMSO neutron PDF data at 100 K. The blue symbols are the total PDF data, the red curve is the best-fit atomic PDF, and the green curve is the fit residual (which contains the mPDF) offset vertically for clarity. (b) Magnetic PDF data (open symbols) and fits (solid curves) at all measured temperatures. Fits described in the main text. (c) schematic of the 3D and 2D combined magnetic order at 100 K.

Fig. 6b shows the magnetic component of the PDF data at each temperature, isolated by subtracting the best-fit atomic PDF calculation from the data at each temperature. To remove any temperature-independent artifacts or background signals from the data, we also subtracted the fit residual at 400 K from all other data sets. Finally, we then added back in the best-fit calculated mPDF at 400 K to all lower-temperature mPDF data sets to avoid removing any real magnetic features from the lower-temperature data. At 100 K, structured features in the mPDF data persist over the full data range shown, indicative of long-range magnetic correlations. The mPDF signal at 100 K can be fit well using a two-component mPDF model that includes the published three-dimensional (3D) G-type magnetic order and a short-range-ordered, purely two-dimensional (2D) component containing the intraplanar AFM correlations [42].

Comparing the weight of the mPDF component to the total PDF signal indicates that the full 5 $\mu_B$ moment expected for $Mn^{2+}$ spins can be accounted for by the combined 3D and 2D components. If just the 3D component is considered, the ordered moment refines to 3.3(1) $\mu_B$, in good agreement with Ref. [42]. The combined 3D and 2D model suggests that only ~3 $\mu_B$ of each local moment participates in the static long-range magnetic order while the rest fluctuates with a short-range intralayer order (as schematically shown in Figure 6c). Within the sensitivity of the mPDF data, no spin canting is observed at 100 K. The moment direction refines to point along the *c* axis for the 3D long-range-ordered component; for the 2D component, the best-fit spin direction is tilted about 45° away from the *c* axis, albeit with a large statistical uncertainty.

For 200 K and above, the mPDF signal is suppressed with increasing inter-spin distance *r*, confirming the presence of short-range magnetic order at these temperatures. The best fits are achieved using the purely 2D model with a finite correlation length that steadily decreases as the temperature increases. At 400 K, only the nearest-neighbor antiparallel magnetic correlation remains meaningful above the noise level in the data. The weight of the mPDF component suggests that at least 1 $\mu_B$ remains antiferromagnetically correlated between nearest-neighbor spins at 400 K, underscoring the robustness of the magnetic fluctuations in this system. As discussed later, Raman scattering at room temperature is consistent with the presence of strong short-range fluctuations.

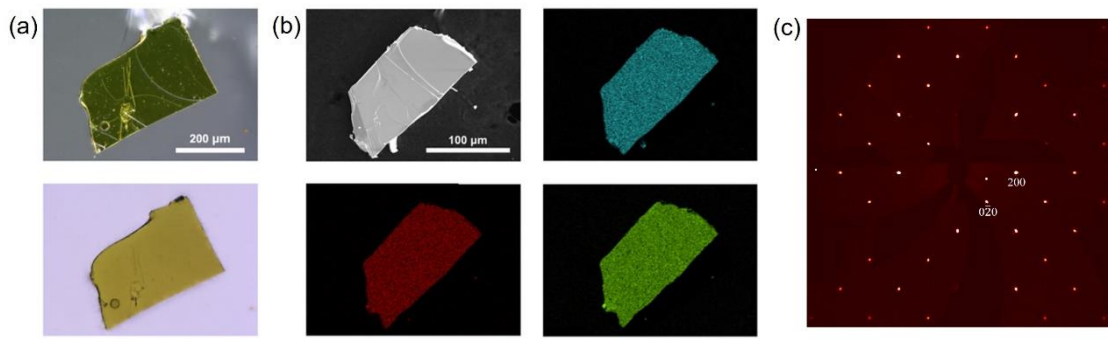

**Figure 7.** (a) Ring-light (up) and transmission-light (below) optical microscope images of a LOMSO crystal. (b) SEM secondary electron image and EDS mapping of a LOMSO crystal. Light blue, red, and green represent La, Mn, and Se, respectively. (c) Measured unsymmetrized (*hk*0) zone of LOMSO using single-crystal X-ray diffraction.

Because the promising properties of altermagnets, especially those of the *d*-wave type, such as the anomalous (thermal) Hall effect and electrical current spin polarization, only arise in a coherent altermagnetic domain, single crystals are needed for future measurements and applications. LOMSO single crystals of tens of μm were previously obtained from polycrystalline pellets but the size was too small for most characterization techniques[62]. We therefore re-attempted crystal growth using a similar high-temperature annealing method, which likely follows a grain growth mechanism (more details in Methods) [63]. The pelletized powder samples post annealing appeared to have had severe evaporation and discoloration on the exposed top surface. Light green and flaky crystals can be found coating the outside of the alumina crucible on the downstream side and were indexed to unit cell dimensions close to those of *α*-MnSe [64]. However, translucent dark green crystals of several hundred micrometers up to sub-mm were retrieved underneath the pellet. They have a thin flaky morphology and occasional step edges on the otherwise smooth surface (see Figure 7a), suggesting a lateral crystal growth mechanism.

To confirm that the crystals obtained are uniform and of the correct phase, we performed a series of analyses. Scanning Electron Microscopy (SEM) imaging of a typical single-crystal plate is shown in Figure 7b. In addition, Energy Dispersive Spectroscopy (EDS) shown in Figure 7b confirms the uniform distribution of elements across the surface of the crystal. The EDS mapping revealed no evidence of elemental segregation or clustering, suggesting that the elements are well-dispersed throughout the sample, with a composition close to the target. The crystal structure was then refined via single-crystal X-ray diffraction (SXRD) at 100 K using an XtaLAB Synergy-R high-flux rotating anode diffractometer. A representative *hk*0 zone is obtained and shown in Figure 7c. The indexed Bragg peaks show the characteristic tetragonal symmetry and systematic absence of Bragg peaks due to the body-centered Bravais lattice. The refinement results are summarized in Table 1–3. Low fitting R factors ($R_1$ = 1.80%, $wR_2$ = 3.61%) are obtained, indicating a good agreement between the structural model and the data. Based on the SXRD data collected, the basal plane of the crystal flake is indexed as (001), which is consistent with the layered nature of the crystal structure normal to [001].

Table 1. Crystal structure refinement data and parameters from SXRD

| Space group type | I4/mmm |
| --- | --- |
| Crystal system | tetragonal |
| Formula (refined) | $La_2Mn_2Se_2O_3$ |
| $a$ (Å) | 4.12740(7) |
| $c$ (Å) | 18.8057(5) |
| $V$ (Å$^3$) | 320.363(14) |
| Z | 2 |
| Temperature | 100(2) K |
| Crystal Size (um) | 100, 70, 20 |
| $\lambda$ | 0.71073 |
| $\mu$ (mm$^{-1}$) | 28.191 |
| Absorption Correction | Gaussian |
| $T_{min}$, $T_{max}$ | 0.165, 0.602 |
| Reflections measured, unique, used | 4028, 526, 526 |
| Resolution ($d_{min}$, $2\theta_{max}$) | 0.442, 107.13 |
| $(hkl)_{max}$ | (-9,8), (-9,9), (-42, 35) |
| Parameters | 15 |
| Method of refinement | Shelx, $|F^2|$ |
| $wR_2$ (all) | 0.0361 |
| $R_1$ (all) | 0.0180 |
| GOF | 1.174 |
| $F_{000}$ | 512 |
| Max residual density | 2.655, -2.519 ( REM Highest difference peak) |
| CSD code | TBD |

Table 2. Atomic sites, isotropic thermal parameters, and occupancies from SXRD

| Atom | Site | $x$ | $y$ | $z$ | $U_{iso}$ (Å$^2$) | Occ. |
| --- | --- | --- | --- | --- | --- | --- |
| La1 | La | 0.5 | 0.5 | -0.18636(2) | 0.00411(3) | 1 |
| Se2 | Se | 0 | 0 | -0.09997(2) | 0.00554(6) | 1 |
| Mn3 | Mn | 0.5 | 0 | 0 | 0.00598(7) | 1 |
| O4 | O | 0 | 0.5 | -0.25 | 0.0056(3) | 1 |
| O5 | O | 0.5 | -0.5 | 0 | 0.0214(12) | 1 |

Table 3. Anisotropic thermal parameters from SXRD

| Atom | $U_{11}$ | $U_{22}$ | $U_{33}$ |
| --- | --- | --- | --- |
| La1 | 0.00280(4) | 0.00280(4) | 0.00673(6) |
| Se2 | 0.00591(8) | 0.00591(8) | 0.0048(2) |
| Mn3 | 0.0044(2) | 0.0031(2) | 0.0104(2) |
| O4 | 0.0040(5) | 0.0040(5) | 0.0089(9) |
| O5 | 0.0033(8) | 0.0033(8) | 0.0058(4) |

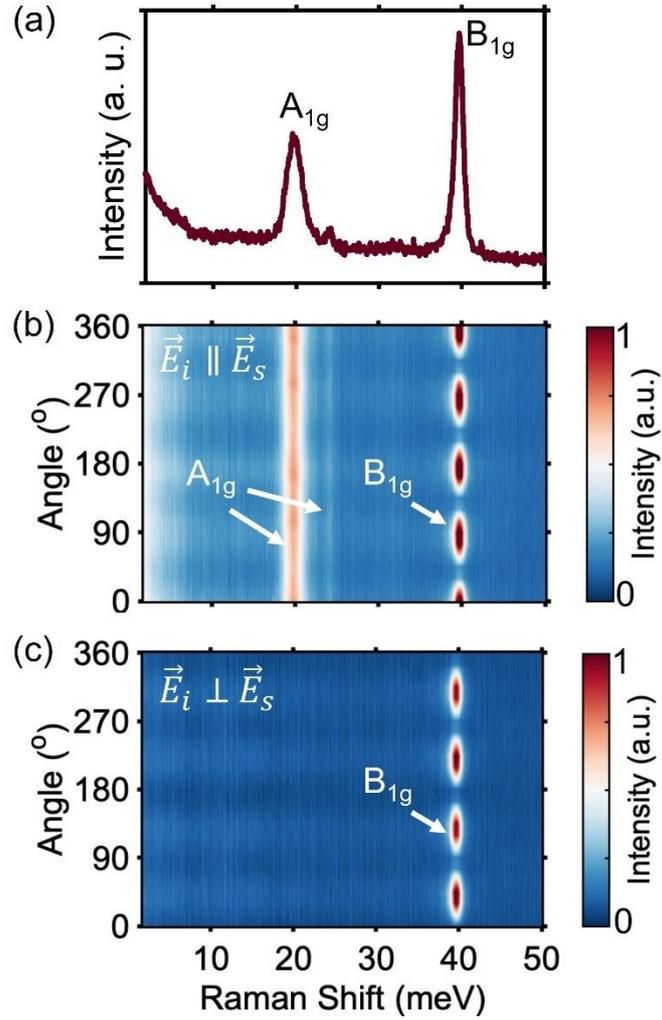

**Figure 8.** (a) Raman spectra of LOMSO at 300 K in parallel (XX) polarization configuration. Color map of the angle-resolved Raman intensity at 300 K in (b) parallel (XX) and (c) crossed (XY) polarization configuration.

To confirm the lattice symmetry and uniformity of the single crystals, we turned to microRaman spectroscopy in backscattering geometry. A typical spectrum recorded with the incident and scattered light polarizations parallel is shown in Fig. 8(a). Similar results (mode position, linewidth, and intensity) were seen in various spots on multiple crystals, indicating their uniformity and high crystal quality. The spectra taken with light propagating along the *c*-axis revealed two strong modes (at 20 meV and 39.7 meV) and one weak mode (at 24.1 meV). Based on the crystal structure and measurement configuration, we expect to observe two modes of $A_{1g}$ symmetry and one mode of $B_{1g}$ symmetry. To confirm this, we measured the angle-resolved Raman response as the light polarization was rotated in the plane. The results for parallel (XX) and crossed-polarization (XY) of the incident and scattered light are shown in Fig. 8 (b) and (c), respectively. As expected for the tetragonal structure with all vertical mirrors, the $A_{1g}$ modes have an angle-independent response in XX configuration but disappear in XY. Similarly, the $B_{1g}$ mode has a four-fold response but with the phase shifted by $45^0$ between XX and XY. Lastly, we note the low-energy tail seen in the spectra is consistent with quasi-elastic scattering from strong magnetic fluctuations often seen in low dimensional and/or frustrated magnetic systems above the ordering temperature [65,66].

As a final note, altermagnets with a *d*-wave spin-momentum texture can host anomalous Hall effect (AHE) as well as other anomalous transport properties [15]. However, whether AHE is allowed depends on the orientation of the Néel vector [67]. LOMSO, which has the moments aligned along the [001] direction will have a zero Hall pseudovector enforced by symmetry. If the moments can be reoriented to within the (001) plane via epitaxial strain, an external magnetic field, or by chemical doping in the transition metal or chalcogen sites (given the competing direct and indirect exchange interactions in the lattice), then a non-zero Hall vector is allowed, and its orientation will depend on the in-plane direction of the moments. LOMSO and its analogues have large gaps which prevent electrical transport properties from being measured. A recent theoretical work has predicted the thermal Hall effect in insulating altermagnets although the effect is still contingent upon the Néel vector orientation [68].

In this context, it is interesting to contrast the magnetic ground state of LOMSO with that of the closely related compounds $RE_2O_3Mn_2Se_2$ with RE = Ce, Pr. According to Ref. [60], while the magnetic space group of the Ce-based compound is the same as LOMSO's (I4′/mm′m), that of the Pr-based compound is different, Im′m′m. A possible explanation for this difference is that the Pr-based compound is still altermagnetic and described by the momentum-spin density $\boldsymbol{S_k} \propto (k_x^2 - k_y^2)\boldsymbol{\sigma}$, but with the magnetic moments ($\boldsymbol{\sigma}$) pointing in-plane rather than out-of-plane. This difference, which is only relevant in the presence of SOC, makes the altermagnetic order parameter of the Pr-based compound transform as the $E_g^-$ irreducible representation of the tetragonal group, rather than $B_{2g}^-$ as in the case of the La- and Ce-based compounds. The $E_g^-$ order parameter not only allows for a non-zero Hall vector and a non-zero ferromagnetic moment, but also causes an orthorhombic lattice distortion, as seen experimentally [60]. Finally, it is also illustrative to compare LOMSO with the related 122-type compound BaMn$_2$As$_2$, which is a member of the ThCr$_2$Si$_2$ (122-type) family that also has space group I4/mmm and displays a G-type magnetic order with out-of-plane moments and wave-vector $\mathbf{k}$ = (0,0,0) [69]. The important difference between LOMSO and BaMn$_2$As$_2$ is that, in the latter, the Mn atoms on the two opposite-spin sublattices are related by inversion. As a result, the magnetic order parameter transforms as the $B_{1u}^-$ irreducible representation, and is thus antiferromagnetic rather than altermagnetic (resulting in magnetic space group I4′/m′m′m). One way to "convert" this antiferromagnetic order into an altermagnetic order would be to apply an out-of-plane electric field (which transforms as the $A_{2u}^+$ irrep), for instance, by growing a very thin film on a substrate, since $B_{2g}^- = B_{1u}^- \times A_{2u}^+$; this is analogous to the strategy proposed in Ref. [70] to obtain an altermagnetic state in FeSe pnictides. The more complex crystal structure of LOMSO, however, automatically provides a proper effective field. This further illustrates the crucial role of the crystalline structure in promoting altermagnetism.

## IV. CONCLUSIONS

In summary, we identified a correlated insulating $d$-wave altermagnet state in La$_2$O$_3$Mn$_2$Se$_2$, which is a member of a large structure family with flexible substitution in the spacer, transition metal as well as the chalcogen positions. The G-type zero wave-vector magnetic ground-state order breaks the parity-time reversal as well as the translation-time reversal symmetries, but is invariant under a combination of time-reversal and 4-fold rotational symmetries, thus giving rise to altermagnetism with a $d_{x^2-y^2}$-wave type spin splitting. The prediction is well aligned with DFT calculations, which show spin splitting along the $\Gamma - X(Y)$, $R - A$ and $X - M$ directions and spin degeneracy along $\Gamma - Z$ and $\Gamma - M$. Polycrystalline samples were synthesized for magnetic and neutron PDF measurements and single crystals of a several hundred micrometer size were successfully obtained through high-temperature annealing. The work calls for future experiments such as (spin-resolved) angle-resolved photoemission spectroscopy to observe the band splitting, although the large gap might bring difficulty unless charge carriers are doped into the system, and thermal Hall measurements if the Néel vector orientation can be adjusted. The layered crystal structure of La$_2$O$_3$Mn$_2$Se$_2$ and the compositional flexibility of its analogues provides a fertile ground for new altermagnet search and optimization in strongly-correlated materials.

**ACKNOWLEDGEMENT**

C.-C.W. and H.J. are supported by an NSF Career grant 2145832. X. L. and F. L. acknowledge financial support from the DOE-BES (No. DE-FG02-04ER46148). Computational resources for this work were supported by CHPC of the University of Utah and the DOE-NERSC. The atomic and magnetic pair distribution function analysis performed by S.R.H. and B.A.F. was supported by the U.S. Department of Energy, Office of Science, Basic Energy Sciences (DOE-BES) through Award No. DE-SC0021134. The neutron scattering experiments used resources at the Spallation Neutron Source, a DOE Office of Science User Facility operated by the Oak Ridge National Laboratory. X.H. and T.T.T. thank the NSF (award NSF-OIA-2227933 and award NSF-DMR-2338014) and the Arnold and Mabel Backman Foundation (2023 BYI grant) for the support. The Air Force Office of Scientific Research supported R.M.F. (phenomenological model) under award number FA9550-21-1-0423 as well as B.S., K.M.K. and K.S.B. (Raman measurements and analysis) under award FA9550-24-1-0110. L.Z. acknowledges the support by the NSF CAREER grant DMR-174774 and Alfred P. Sloan Foundation.

# Supplementary Information

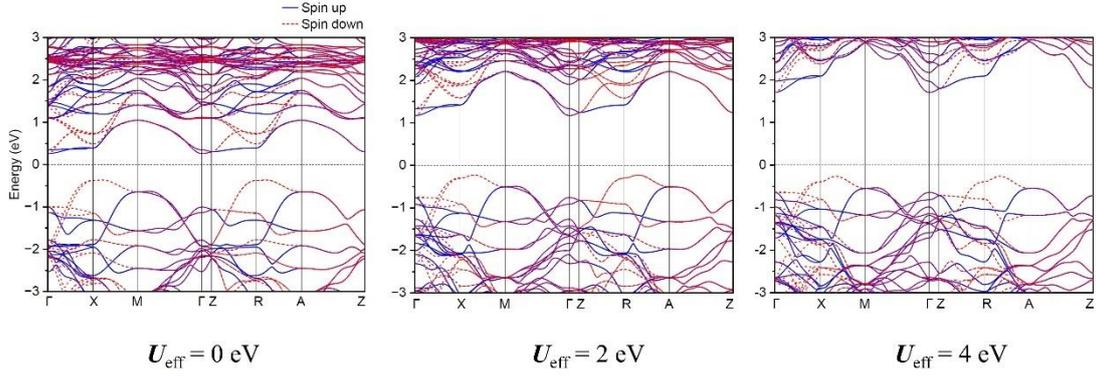

**Figure S1.** Spin-projected band structure of LOMSO assuming a G-type compensated magnetic state without SOC under various $U_{eff}$'s.

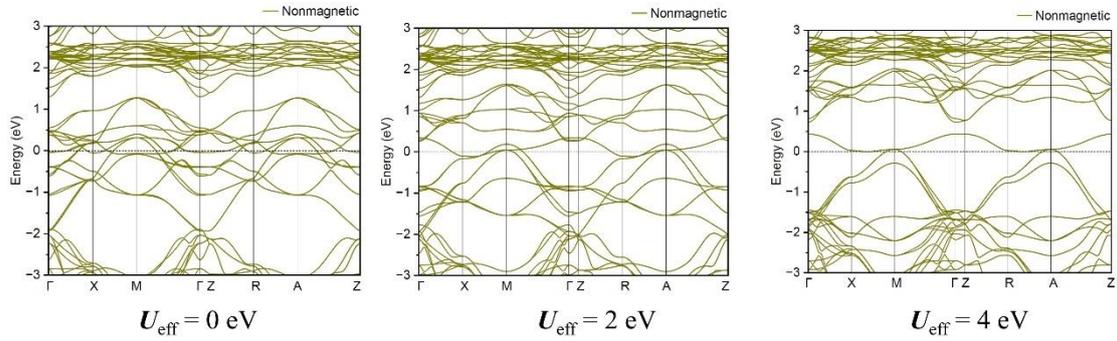

**Figure S2.** Band structure of LOMSO assuming a non-magnetic state without SOC under various $U_{eff}$'s.

**Table S1.** Computed energies of different magnetic states using various functionals and $U_{eff}$'s.

| $U_{eff}$ (eV) | Relative energy (eV/Mn atom) | | | | | |
|---|---|---|---|---|---|---|
| | GGA+U | | | LDA+U | | |
| | Compensated (G-type) | FM | NM | Compensated (G-type) | FM | NM |
| 0 | 0 | 0.241 | 2.024 | 0 | 0.232 | 1.544 |
| 2 | 0 | 0.163 | 3.152 | 0 | 0.171 | 2.817 |
| 4 | 0 | 0.117 | 3.973 | 0 | 0.114 | 3.686 |